%% file: main.tex
\documentclass{article}
\usepackage{siunitx}
\usepackage[left=2cm, right=2cm, top=1.5cm, bottom = 1.5cm]{geometry}
\usepackage{amssymb, amsmath}
\usepackage{dsfont}  
\usepackage[version=4]{mhchem}
\usepackage{chemfig}
\usepackage{multirow}
\usepackage{csquotes}
\usetikzlibrary{calc}
\usepackage{float}

\input{macros}

\title{Virtual materials testing of ASSB cathodes combining AI-based stochastic 3D modeling and numerical simulations}
\author{A. Dufter$^{1*}$, S. Weber$^1$, O. Furat$^2$, J. Schubert$^3$, R. Rekers$^3$, M. Luczak$^4$, E. Glatt$^4$, A. Wiegmann$^4$, \\A. Bielefeld$^3$ and V. Schmidt$^1$}
\date{}

\begin{document}

\maketitle
\begin{center}
	\it
	$^1$Institute of Stochastics, Ulm University, Helmholtzstraße 18, 89069 Ulm, Germany
	\\
    $^2$Applied AI and Data Science Unit, University of Southern Denmark, DK-5230 Odense, Denmark
    \\
	$^3$Center for Materials Research (ZfM), Justus Liebig University Giessen, Heinrich-Buff-Ring 16, Giessen 35392, Germany
	\\
	$^4$Math2Market GmbH, Richard-Wagner-Straße 1, 67655 Kaiserslautern, Germany
\end{center}
$^*$Corresponding author. \textit{Email address:} anina.dufter@uni-ulm.de (Anina Dufter)\\

\begin{abstract}
The performance of all-solid-state battery (ASSB) cathodes strongly depends on their microstructure. Optimizing the cathode morphology can therefore enhance effective macroscopic properties such as ionic and electronic conductivity. The search for optimized microstructures can be facilitated by virtual materials testing:
By integrating image analysis and stochastic microstructure modeling to generate a wide range of realistic 3D microstructures and evaluate their effective macroscopic properties by means of numerical simulations, thereby reducing the need for extensive physical experiments.
This approach allows for the investigation of structure-property relationships through parametric regression models that incorporate relevant geometrical descriptors of microstructures such as volume fractions, mean geodesic tortuosities, specific surface areas, and constrictivities. 
By linking these geometrical descriptors to macroscopic properties, virtual materials testing provides quantitative insight into how microstructure influences material performance.
In the present paper, this framework is applied for ASSB cathodes. 
In addition, by systematically varying model parameters, a broad range of 3D microstructures can be generated, which remain close to the original cathode morphology while inducing targeted changes in selected geometrical descriptors.
The resulting database enables the calibration of regression models whose predictive performance is assessed by comparing predicted and simulated effective properties such as the ionic and electronic conductivity, thereby quantifying how accurately combinations of geometrical descriptors can explain and predict variations in effective macroscopic properties.

\end{abstract}

structure-property-relationship, conductivity, virtual materials testing.

\section{Introduction}
Lithium-ion batteries are extensively employed due to their high energy density, low mass, and low self-discharge rate, and have increasingly become a key technology in electromobility~\cite{korthauer2018lithium, passerini2020batteries}.
Nevertheless, the reliance on flammable liquid electrolytes raises significant safety concerns, while challenges such as dendrite formation, electrolyte leakage, and intrinsic limitations in capacity and energy density remain~\cite{aziam2022solid}.
All-solid-state batteries (ASSBs), in which liquid electrolytes are replaced by solid-state counterparts, present a promising alternative by offering enhanced safety, superior thermal stability, and extended service life~\cite{aziam2022solid, chen2021research, lim2020review}.
However, challenges such as insufficient ionic conductivity and high interfacial resistance remain to be addressed~\cite{janek2023challenges, ren2023oxide, zhang2018synthesis}.

These limitations are strongly influenced by the microstructure of the battery components, particularly, the microstructure of the electrodes and the interfaces between phases such as the active material and solid electrolyte.
In this context, the microstructure of materials, which can be characterized using advanced imaging techniques, plays a crucial role in determining their effective properties and is therefore a key focus of ongoing research~\cite{asheri2024microstructure, vu2023towards, marmet2024multiscale, clausnitzer2023optimizing}.
For ASSB cathodes this has been extensively studied in~\cite{park2020digital}.
To characterize such microstructures, commonly employed imaging methods include computed electron tomography, computed X-ray tomography---such as nano-CT and micro-CT---and focused ion beam scanning electron microscopy (FIB-SEM)~\cite{liu2025fib, carazo2015three, maire2014quantitative}.
These techniques enable high-resolution two- or three-dimensional reconstructions of the underlying microstructure, allowing the computation of geometrical descriptors of the microstructure such as volume fractions, surface areas, tortuosities, and constrictivities~\cite{ohser2000statistical}.

Since the geometry of the microstructure significantly affects effective macroscopic properties, understanding the relationships between microstructure and macroscopic properties is of considerable interest~\cite{stenzel2017big, prifling2021large, neumann2020quantifying}.
Optimizing material properties typically requires exploring a vast space of material combinations and microstructures. In the context of ASSB cathodes, relevant properties include ionic and electronic transport, mechanical contact, and interfacial stability~\cite{nam2025spotlighting}.
Conducting such investigations purely through experimental testing can be extremely time-consuming and costly.
Therefore, virtual ASSB microstructures offer a valuable alternative and play a crucial role in ASSB research~\cite{park2020digital, kim2023synergistic, kim2020diffusion}.  
The concept of virtual material testing~\cite{virtualMaterialTesting} is built on such virtual microstructures, enabling the systematic analysis of structure-property relationships without the high costs associated with large production volumes and imaging of samples manufactured under varying process conditions~\cite{neumann2020quantifying}.
In contrast to a fully experimental approach, only a small number of samples is required.
These are used to calibrate a parametric stochastic 3D model based on methods from stochastic geometry~\cite{jeziorski2024stochastic, theodon2024vox}.
Once calibrated, the model can be regarded as a digital twin and allows for the generation of virtual, but realistic structures that are ideally statistically equivalent to the experimentally measured counterparts~\cite{jeziorski2024stochastic,furat2021artificial, neumann2023data}.
To quantitatively validate the model, geometrical descriptors such as the volume fraction or the specific surface area are computed for the tomographic image data and
compared those computed for the model realizations~\cite{neumann2019pluri, marmet2023stochastic}.

Based on a database of virtual yet physically realistic microstructures, effective macroscopic properties---specifically ionic and electronic conductivity in ASSBs---can be computed by means of numerical simulations~\cite{prifling2021large,neumann2020quantifying}.
Such numerical simulations have been employed, for example in~\cite{inoue2017numerical}, where effective electronic and ionic conductivities in lithium-ion batteries were investigated through both numerical and experimental analysis, examining the relationship between porous electrode structure and effective transport properties.
Numerical simulations of effective properties form the basis for predictive models that relate geometrical descriptors to macroscopic transport behavior~\cite{prifling2021large, fohst2022influence, daubner2024microstructure}.
To this end, regression-based prediction formulas~\cite{fohst2022influence, prifling2019parametric, stenzel2016predicting} incorporating multiple geometrical descriptors are employed to systematically investigate the influence of microstructure on effective properties.
Previous studies have shown that several geometrical descriptors such as volume fraction, tortuosity and constrictivity are well suited for this purpose~\cite{stenzel2017big,prifling2021large}.

Building on this idea and aiming to extend this concept to ASSB microstructures, we apply an approach that combines so-called generative adversarial networks (GANs)~\cite{goodfellow2014generative} with a slightly simplified version of a stochastic geometry model introduced in~\cite{furat2025generative}.
This approach overcomes the limitations of GANs---namely their restricted capability for systematic parameter variation due to numerous and often uninterpretable model parameters (i.e., the set of trainable weights)---as well as those of stochastic geometry models, which may struggle to capture highly complex morphologies.
This model has been fitted in~\cite{furat2025generative} to three experimentally measured ASSB cathodes composed of identical constituent materials, differing only in the preparation of the solid electrolyte (SE), specifically the milling media used to produce different SE particle systems~\cite{anjaData}.
These fitted models form the foundation for systematic parameter variation and the generation of a diverse database of virtual 3D realizations.
In stochastic models, the volume fraction can often be controlled directly, allowing for the generation of microstructures that exhibit approximately uniformly distributed volume fractions~\cite{prifling2021large, marmet2023stochastic}.
However, selecting model parameters such that the generated microstructures attain prescribed values of certain geometrical descriptors---particularly those closely related to transport processes, such as geodesic tortuosity and constrictivity---remains challenging.
This limitation applies in particular to more complex stochastic 3D models, such as the one employed in the present work.
To address this challenge while still ensuring that the resulting scenarios are within the scope of realistic ASSB cathode microstructures, a two-step gradient-based approach is introduced, that, to the best of our knowledge, has not been previously applied in this context. 
Subsequently, regression-based prediction formulas---which have not been widely investigated for ASSBs---can be explored after computing ionic and electronic conductivities.

These structure-property relationships can provide the foundation for inverse microstructure design, which seeks to determine geometrical descriptors required to achieve targeted effective macroscopic properties, such as high ionic conductivity, sufficient electronic conductivity, and low tortuosity.
Rather than predicting transport properties from a given microstructure, the relationship is inverted and formulated as an optimization problem to identify optimal geometrical descriptor ranges and model parameters that yield the desired effective macroscopic properties.
Inverse materials design has been applied, for instance, in~\cite{zou2025inverse}, which introduces a framework for the development of lithium metal battery architectures by combining high-throughput phase-field simulations with machine-learning-based optimization techniques.
Through a systematic exploration of electrode and separator design variables, the study identifies the fundamental mechanisms influencing dendrite growth and their impact on overall cell performance.
In this context, the present work contributes by establishing robust quantitative structure-property relationships, thereby providing a necessary basis for future inverse microstructure design approaches.

The paper is structured as follows. First, a small introduction is given on the materials used in the construction of the three ASSB cathodes, along with a description of the stochastic geometry model employed in this study. 
Subsequently, several geometrical descriptors relevant to effective conductivity are discussed, and the regression models used to analyze structure-property relationships are introduced.
Additionally, a methodology for generating the large database of virtual ASSB cathodes is presented that combines interpolation and gradient-based approaches.
The results obtained from these methods are then presented and discussed.
Finally, the paper concludes with a summary of the main findings.

\section{Materials and methods}
The following section first provides a brief overview of the investigated cathode materials and the corresponding image data acquisition procedures. 
Next, we introduce the stochastic 3D model and the geometrical descriptors used to characterize microstructures obtained from the tomographic imaging.
Subsequently, the effective macroscopic properties---namely ionic and electronic conductivities---are presented together with regression models that relate geometrical descriptors to the effective properties and enable the analysis of transport properties.
Finally, an approach combining parameter interpolation and gradient-based methods is described, which facilitates the generation of a large database of virtual, but realistic ASSB cathode microstructures.
\subsection{Description of material and data acquisition} \label{subsec: material and data acquisition}
In this section, we briefly present the ASSB cathode materials used for the acquisition of experimentally measured image data.
The ASSB cathodes considered in this paper consist of three phases, namely the solid electrolyte (SE), the active material (AM), and the pore space (P).
Specifically, glassy \ce{Li_3PS_4-0.5LiI} is used as the SE and \ce{LiNi_{0.83}, while Mn_{0.06}Co_{0.11}O_2} serves as the AM~\cite{anjaData}.
The three data sets considered in this work were obtained by varying the milling media in a planetary ball mill (Pulverisette 7, Fritsch, Germany) to produce different SE particle systems.
The milling media diameters were chosen as $\SI{1}{\milli\meter}$, $\SI{3}{\milli\meter}$ and $\SI{10}{\milli\meter}$.
The resulting data sets from combining these SE  particle systems with the AM are referred to as BM01, BM03 and BM10, respectively.

The microstructure of these three ASSB cathodes have been measured using plasma focused ion beam scanning electron microscopy (PFIB-SEM), which is an advanced imaging and sample preparation technique.
PFIB-SEM combines plasma ion beam milling (PFIB), which precisely removes material to expose internal structures, with scanning electron microscopy (SEM) providing high-resolution images of the sample surface.
This approach enables precise cross-sectioning and 3D reconstruction of complex materials allowing accurate characterization of microstructural features.
The resulting 3D images of the ASSB cathode microstructures have a resolution of $\SI{0.1}{\micro\meter}$ and have been segmented into the three phases, namely the SE and AM as well as the remaining pore space.
For computational purposes, the images are rescaled to a voxel size of $\SI{0.2}{\micro\meter}$.
We refer to~\cite{anjaData} for more details regarding material composition and imaging.

Note that during fabrication, ASSB cathode powders are pressed uniaxially into cylindrical half-cells, without binders or conductive additives.
As a result, microstructural characteristics may depend on direction, so the cathode cannot be assumed to be isotropic in 3D.
For the ASSB cathodes studied, the pressing direction aligns with the $z-$axis, and inspection of 2D sections shows that the structure differs between planes, indicating anisotropy along $z$.
However, sections parallel to the $x$-$y$ plane appear statistically similar under rotation around $z$, suggesting cylindrical isotropy to be a reasonable approximation in-plane.
See~\cite{furat2025generative} for more details on the investigation of anisotropy with respect to the data sets BM01, BM03 and BM10.

\begin{figure}[H]
    \centering
    \includegraphics[width=0.9\linewidth]{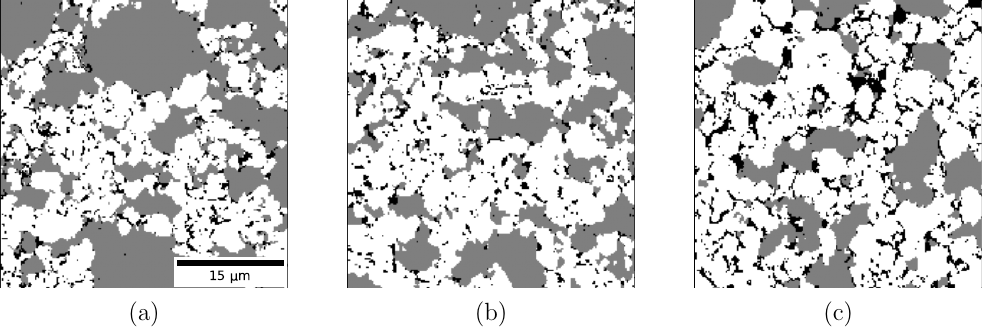}
    \caption{Exemplary chosen 2D sections of the experimentally measured 3D images of BM01 (a), BM03 (b) and BM10 (c). The AM, the SE and the pore space are represented in white, gray and black, respectively.}
    \label{fig:placeholder}
\end{figure}
 
\subsection{Model description} \label{subsec: model description}
In this section, the framework for modeling the ASSB cathode microstructure is presented. In the present paper, we employ the parametric stochastic 3D modeling approach introduced in~\cite{furat2025generative}, with the simplification that the model parameter $\gamma$, controlling the degree of correlation between the random fields, is fixed at $0$. This is consistent with the calibrated parameters of the three experimentally acquired image data sets. 
This modified model is described by a parameter vector $\theta \in \Theta$, where $\Theta= (0, \infty)^4\times\R^2$ denotes the set of valid parameter vectors.
It is based on random fields $X=\{X(t): t\in \Z ^3\}$, $Y=\{Y(t):t\in \Z ^3\}$ given by 
\begin{align}
    X=X'+X'' \quad \text{and} \quad Y=Y'+Y'',
\end{align}
where $X''$ and $Y''$ are two normalized independent and motion invariant (stationary and isotropic) GRFs that are independent of two motion invariant $\chi ^2$-fields $X', Y'$ with $n$ degrees of freedom. 
The latter two are given by
\begin{align}
    X' = \sum _{j=1}^n X_{c,j}^2 \quad \text{and} \quad Y' = \sum _{j=1}^n Y_{c,j}^2,
\end{align}
where $X_{c,j}, Y_{c,j}, j=1, \dots, n$ are independent copies of two normalized independent GRFs $\tilde{X}$ and $\tilde{Y}$, respectively.
Each of these normalized, motion invariant GRFs, i.e. $\Tilde{X}, \Tilde{Y}, X''$ and $ Y''$, can be uniquely characterized by its covariance function.
For each of these GRFs we choose the same parametric family of functions to describe their underlying covariance function $\rho_\alpha: \Z^3 \to [0,1]$, namely,
\begin{align}
    \rho _\alpha (h) = e^{-(\alpha h)^2},
\end{align}
where $\alpha \in (0,\infty)$ is a model parameter, which we refer to as the covariance parameter.
From here on, we denote the covariance parameters of the GRFs $\Tilde{X}, \Tilde{Y}, X''$, and $ Y''$  by $\alpha_{\Tilde{X}}, \alpha_{\Tilde{Y}}, \alpha_{X''}$, and $\alpha_{Y''}$, respectively.
A relatively simple covariance function is intentionally selected, as it facilitates  systematic parameter variation in the parameter study.

On a sampling window $W = \{1, \dots, n_x\} \times \{1, \dots, n_y\} \times \{1, \dots, n_Z\} \subset \mathbb{Z}^3$, where $n_x, n_y, n_z > 0$ denote the the side lengths, a voxelized 3D microstructure can be generated with the stochastic 3D model. More precisely, the stochastic model is given by $\Xi \colon W \times \{1,2,3\} \to \{0,1\}$, such that $\Xi  (t,1), \Xi  (t,2)$ and $ \Xi  (t,3)$ are representing the SE, AM and the pore space respectively. Formally, it applies
\begin{align}
    \excOne &= \left\{
	\begin{array}{ll}
		1, & \text{if }  X(t) \geq \lambda _X,\\
		0, & \text{else,}
	\end{array}
	\right.\\
    \excTwo &= \left\{
	\begin{array}{ll}
		1, & \text{if }  \excOne=0 \text{ and }Y(t) \geq \lambda _Y,\\
		0, & \text{else,}
	\end{array}
    \right.\\
    \excThree &= 1-\excOne -\excTwo ,
\end{align}
where $\lambda _X, \lambda _Y\in \mathbb{R}$ are the threshold parameters.
Thus, voxels $t\in \Z ^d$ with $\excOne = 1$ are associated with the SE, those with $\excTwo = 1$ with the AM, and those with $\excThree=1$ with the pore space.
Note that the stochastic model comprises six model parameters in total, i.e., the two threshold parameters $\lambda _X$ and $ \lambda _Y$, as well as the four parameters from the covariance functions of the normalized stationary GRFs $\alpha_{\Tilde{X}}, \alpha_{\Tilde{Y}}, \alpha_{X''}$, and $\alpha_{Y''}$.

Since this stochastic 3D model is parametric, we refer to $\Xi$ with a specification for its model parameter vector $\theta \in \Theta$ by $\Xi _\theta$.
A valid parameter vector is then given by 
\begin{align}
    \theta = (\alpha_{\Tilde{X}}, \alpha_{\Tilde{Y}}, \alpha_{X''}, \alpha_{Y''},\lambda _X, \lambda _Y) \in \Theta.
    \label{eq: parametervector}
\end{align}
Due to the lack of analytical formulas that would allow for a direct calibration to the image data, GANs are employed to perform the model calibration.
For more details on the construction and calibration of the model we refer to \cite{furat2025generative}.

Note that some small anisotropy effects can be observed in the experimentally measured data sets BM01, BM03, and BM10 with respect to the $z$-direction.
To account for this, anisotropy is introduced by scaling the originally isotropic model $\Xi$ along the $z$-axis, see~\cite{furat2025generative}.
The corresponding scaling parameter is denoted by $\aniso \in \R$.

\subsection{Geometrical descriptors and their estimation} \label{subsec: morphological descriptors and their estimation}
The geometrical descriptors presented in this section are defined for 3D image data. 
Since $\Xi$ is a stochastic model, these descriptors are computed for realizations of $\Xi$.
\paragraph{Volume fraction}
One of the most important geometrical descriptors is the volume fraction $\varepsilon \in [0,1]$ of a phase, which quantifies the proportion of this to the total volume of the 3D image. 
The volume fractions can be estimated using the so called point-count method~\cite{StGeoAppl}, where the number of voxels belonging to the phase under consideration is divided by the total voxel count.
In the following, geometrical descriptors are denoted by symbols that include a subscript indicating the phase under consideration.
The subscripts AM, SE, and P correspond to the active material, solid electrolyte, and pore phase, respectively.
For instance, $\varepsilon _\AM$ represents the volume fraction of the active material, while $\varepsilon _\SE$ and $\varepsilon _\PS$ denote the volume fraction of the solid electrolyte and pore space.

\paragraph{Specific surface area}
Another relevant geometrical descriptor is the specific surface area $S$, which describes the mean surface area per unit volume.
For a given phase, it is computed as the total interfacial area between this phase and all remaining phases.
For a 3D image, it can be estimated using an approach presented in~\cite{sfa_schladitz2006} that is based on convolution of the image with a $2\times 2\times 2$ mask.

\paragraph{Geodesic tortuosity}
Given a certain direction of transport, the geodesic tortuosity, denoted by $\tau \geq 1$, is a measure for shortest paths in relation to the thickness of the material and a crucial geometrical descriptor regarding virtual material testing due to its influence on effective macroscopic properties such as diffusivity or permeability~\cite{prifling2021large,virtualMaterialTesting, stenzel2017big}. 
It quantifies the length of pathways connecting a predefined starting plane to a target plane, restricted to voxels belonging to the phase of interest, relative to the distance between the two planes, see~\cite{TortConstr} for a formal definition. In this paper, the geodesic tortuosity is evaluated in $z$-direction. 
The descriptor is estimated from image data by using the Dijkstra algorithm~\cite{dijkstra_jungnickel2005graphs} to compute the shortest path length between the starting and target plane. 
The paths are normalized by dividing them by the distance between both planes. Applying this to all points in the starting plane within the considered phase leads to a distribution of tortuosity values. 
For each phase $p\in \{\AM,\SE,\PS\}$, the descriptors $\mu (\tau_p)$ and $\sigma (\tau_p)$ denote the mean and standard deviation across all shortest paths computed in the given direction.

\paragraph{Constrictivity}
To quantify bottleneck effects in porous media, the constrictivity $\beta \in [0,1]$ is introduced.
It can be computed by using two concepts, namely the simulated mercury intrusion porosimetry (SMIP) and the continuous phase size distribution (CPSD). 
The latter one is a function $\mathrm{CPSD}\colon[0, \infty) \to [0,1]$ characterizing the size of bulges of the phase under consideration. 
More precisely, for a given radius $r\geq 0$, the value $\mathrm{CPSD}(r)$ corresponds to the volume fraction of the phase of interest that can be covered by (overlapping) balls of radius $r$, which do not intersect any other phase.
Furthermore, $r_{\mathrm{max}}$ denotes the radius of the characteristic bulge, i.e. the largest radius for which the continuous phase size distribution is greater than or equal to half of the volume fraction of the considered phase.
The simulated mercury intrusion porosimetry $\mathrm{SMIP}\colon [0, \infty) \to [0,1]$ corresponds to the volume fraction that can be covered by a ball of radius $r$ that forms an intrusion from a predefined direction.
Similarly to $r_{\mathrm{max}}$, we denote the radius of the characteristic bottleneck, i.e. the largest radius for which the normalized mercury intrusion porosimetry is at least half of the volume fraction of the considered phase, by $r_\mathrm{min}$. 
Both quantities can be computed using morphological openings and Euclidean distance transform~\cite{soille1999morphological, munch2008contradicting}.
Moreover, note that, since SMIP, unlike CPSD, is dependent on a predefined direction, it holds that $\mathrm{SMIP}(r) \leq \mathrm{CPSD(r)}$ for all $r\geq 0$.
Then, the constrictivity is given by $\beta = (\frac{r_\mathrm{min}}{r_\mathrm{max}})^2$, where values close to zero indicate strong bottleneck effects, while values close to one indicate almost no restrictions.
We refer to~\cite{TortConstr} for a formal definition.

\subsection{Parameter study}\label{subsec: sim study}
Since, as in many practical scenarios, the acquisition of real microstructure data is often limited in both quantity and variability---due to the high costs and time constraints associated with production and imaging---virtual materials testing offers an effective alternative.
Virtual but realistic microstructures are simulated, and their effective macroscopic properties are evaluated through numerical simulations.
An important prerequisite for building an extensive database is the availability of a parametric stochastic model, as introduced in Section~\ref{subsec: model description}, which enables a systematic variation of experimentally derived parameter vectors to generate microstructures that exhibit new, previously unobserved statistical characteristics.

Based on the three experimentally measured data sets BM01, BM03, and BM10 presented in Section~\ref{subsec: material and data acquisition}, the goal is to construct a large database of ASSB cathode microstructures that is as diverse as possible with regard to geometrical descriptors, yet reasonably realistic. 
To achieve this, we proceed in two stages.
First, parameter vectors associated with the three different data sets are interpolated, enabling the generation of virtual microstructures that represent intermediate states between the experimentally observed microstructures. However, the aim is to explore a wider range of virtual but realistic microstructures. 
Therefore, we consider the gradient of a function that maps a parameter vector to the mean geometrical descriptor computed over five independent model realizations.
This gradient indicates how changes in the parameters affect the descriptor, enabling  purposeful adjustments to the microstructure.
As this gradient cannot be computed analytically, it is approximated using difference quotients.
Applying this methodology to all interpolated parameter vectors enables a targeted and systematic exploration of the microstructure space beyond purely interpolated states, thereby substantially enlarging the accessible data space.

\begin{figure}[H]
    \centering
    \scalebox{0.7}{
    \includegraphics[width=0.99\textwidth]{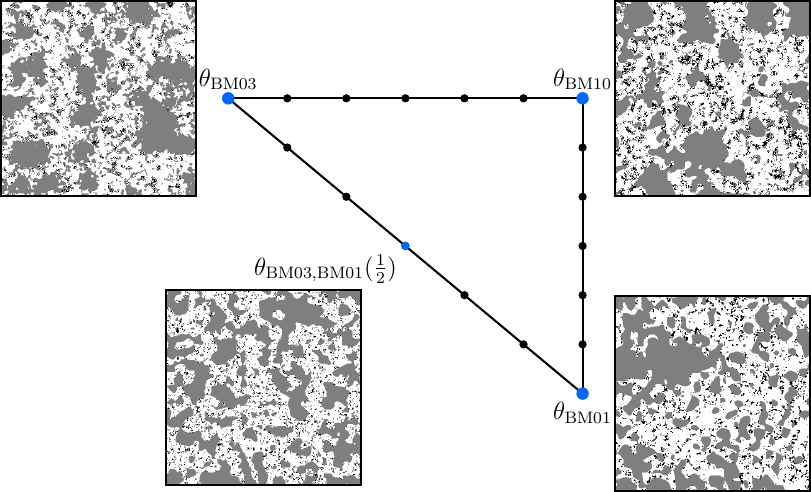}
    }
    \caption{Schematic illustration of the interpolation procedure between the model parameter vectors calibrated to the data sets BM01, BM03 and BM10. Exemplary 2D slices of 3D model realizations of $\theta _{\text{BM01}},\theta _{\text{BM03}},\theta _{\text{BM10}}$ and $\theta _{\text{BM03,BM01}}(\frac{1}{2})$ are shown.}
    \label{fig: interpolation scheme}
\end{figure}

We start by introducing the interpolation procedure to generate virtual microstructures.
To this end, for two data sets $d_1, d_2\in \{\mathrm{BM01, BM03, BM10}\}$ with $ d_1\neq d_2$, we select parameter vectors located on the connecting line between $\theta _{d_1}$ and $\theta _{d_2}$, where $\theta _{d_1}, \theta _{d_2} \in \Theta$ are the parameter vectors of the stochastic model calibrated to $d_1$ and $d_2$. Note that $\Theta$ denotes the space of admissible parameter vectors, defined as in Section~\ref{subsec: model description}. 
For the interpolation between two parameter vectors, we consider the mapping $\theta _{d_1,d_2}\colon (0,1) \to \Theta$ given by
\begin{align} \label{equ: straight line}
    \theta_{d_1,d_2} (t) = \theta _{d_1} +  t (\theta_{d_2} - \theta_{d_1}) \in \Theta, 
\end{align}
for $t \in (0,1)$, \ $d_1, d_2\in\{\mathrm{BM01, BM03, BM10}\}$ and $d_1\neq d_2$.
A schematic visualization of this interpolation approach is shown in Figure~\ref{fig: interpolation scheme}.
Recall that some small anisotropy effects can be observed in the experimentally measured data sets with respect to the $z$-direction.
The scaling factor introduced in Section~\ref{subsec: model description}, denoted by $\aniso \in \R$, is also interpolated analogously to the parameter vector.
Specifically, this yields $\aniso _{d_1, d_2} = \aniso _{d_1} + t (\aniso _{d_2} -\aniso _{d_1})$ for $t\in(0,1)$, $d_1, d_2\in \{\mathrm{BM01, BM03, BM10}\}$, $d_1\neq d_2$.
For the parameter study in this paper, the interpolation parameter $t$ is chosen such that five equidistant points are obtained between each pair of parameter vectors $\theta _{d_1}$ and $\theta _{d_2}$, i.e., $t\in\{\frac{1}{6}, \frac{2}{6},\dots , \frac{5}{6} \}$.
For each interpolated parameter vector, a corresponding model realization is generated, resulting in a 3D microstructure of size $\SI{80.2}{\micro\meter} \times \SI{80.2}{\micro\meter} \times \SI{80.2}{\micro\meter}$, corresponding to a $401\times 401 \times 401$ voxel grid.
In order to systematically explore further structural scenarios, we consider a gradient-based approach.
While this method can be applied to various geometrical descriptors, for simplicity, the concept is only introduced for the specific surface area of the AM. 
More precisely, the function which maps the parameter vector of the stochastic model to the specific surface area of the AM is denoted by  $f_{S _\AM}\colon\Theta \to [0,\infty )$.
However, the same procedure applies analogously for $S_\text{SE}, \varepsilon _p, \mu (\tau _p), \beta _p$ and the corresponding functions $f_{S _\SE}\colon\Theta \to [0,\infty )$ as well as $f_{\varepsilon _p} \colon\Theta \to [0,1]$, $f_{\mu (\tau _p)}\colon \Theta \to [1, \infty)$, and $f_{\beta _p}\colon \Theta \to [0,1]$ for $p\in\{\AM, \SE\}$ that maps the parameter vector of the stochastic model on the corresponding geometrical descriptor.
However, due to the stochastic variation between model realizations for the same parameter vector, we consider the function $\sfaAM \colon\Theta \to [0,\infty)$, which maps a parameter vector $\theta \in \Theta$ to the mean specific surface area of the AM evaluated over five realizations of the model corresponding to $\theta$, thereby reducing the influence of statistical fluctuations and potential outliers. 
For simplicity, the methodology is illustrated for the connecting line between $d_1$ and $d_2$.
The same applies analogously to all other connecting lines.
We denote $\theta _ {d_1, d_2}(\frac{1}{6}), \theta _ {d_1, d_2}(\frac{2}{6}), \dots, \theta _ {d_1, d_2}(\frac{5}{6})$ by $\theta _1, \theta _2, \dots \theta _5$, respectively.
To generate model realizations that exhibit significant variations in the specific surface area, we approximate the directional derivative of $\sfaAM$ at $\theta _k$ by means of a difference quotient.
In particular, we use a small perturbation $\delta >0$ in the direction of the standard basis vector $e_q\in \mathbb{R}^m$ to estimate the gradient $(\nabla \sfaAM)(\theta _k)$ by
\begin{align} \label{equ: gradient approach}
    (\nabla \sfaAM)(\theta _k) = \left (
    \frac{\sfaAM (\theta _k + \delta \cdot e_q) - \sfaAM(\theta_k)}{\delta}
    \right ) _{q=1} ^m,
\end{align}
where $m\in \mathbb{N}$ is the dimension of the parameter space. 
This (approximation of the) gradient provides the direction of the steepest change in $\sfaAM$ at each $\theta _k$ and allows for a systematic generation of virtual microstructures in a neighborhood of $\theta _k$.
The resulting parameter vectors are of the form 
\begin{align} \label{equ: new gradient parameter vectors}
    \Tilde{\theta}_k(h) = \theta _k + h \cdot \nabla \sfaAM(\theta _k),
\end{align}
where the scalar $h \in \mathbb{R}$ controls the step size along the gradient direction.
Note that for all model realizations generated within the gradient-based approach necessary for the evaluation of $\sfaAM$---namely for the parameter vectors  $\theta _k$ and $\theta_k + \delta \cdot e_q$, see Eq.~\eqref{equ: gradient approach}, as well as the parameter vector $\tilde{\theta} _k$ defined in Eq.~\eqref{equ: new gradient parameter vectors}---the anisotropy value $\aniso$ is fixed to $\aniso _k$, i.e., to the anisotropy factor of the respective base parameter vector $\theta _k$.
The anisotropy is fixed in this way because the values obtained from the three experimentally measured datasets are very similar.
More precisely, the scaling factors corresponding to BM01, BM03, and BM10 are $0.92,\, 0.94$, and $1$, respectively.

\begin{figure}[H]
    \centering
    \includegraphics[width=0.8\textwidth]{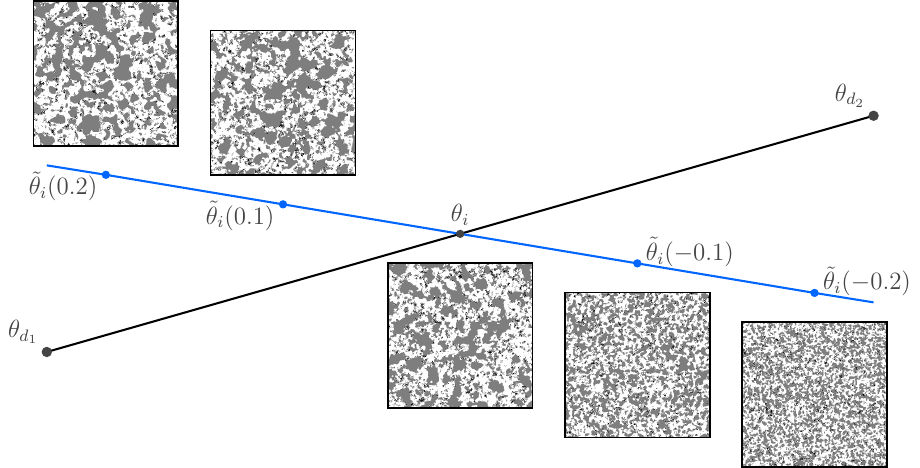}
    \caption{Schematic illustration of the gradient-based approach illustrated for the specific surface area. The blue line corresponds to the gradient.}
    \label{fig: gradient approach schmee}
\end{figure}

This methodology can be similarly extended to the specific surface area of the SE, as well as to additional geometrical descriptors, including volume fraction, mean geodesic tortuosity, and constrictivity.
A schematic illustration of the gradient-based approach is shown in Figure~\ref{fig: gradient approach schmee}.
Overall, this approach provides a framework for systematically generating microstructural variations around a given parameter vector $\theta _k$, allowing for a broader, but well-controlled exploration of the parameter space.
The incremental parameter shift $\delta$ is set to $0.05$.
When computing the gradient with respect to the volume fractions, mean geodesic tortuosity, and specific surface area, the step size $h$ is successively chosen from the set $\{-0.2, -0.1, 0.1, 0.2\}$.
For the constrictivity, smaller values $\{-0.05, -0.025, 0.025, 0.05\}$ are used, as larger step sizes lead to unrealistic cathode microstructures.
To further ensure physical plausibility, the admissible range of $\alpha _r$ is restricted to $[0.05, \infty)$ for all GRFs $r\in \{\Tilde{X}, \Tilde{Y}, X'', Y''\}$ introduced in Section~\ref{subsec: model description}.
For each resulting parameter $\Tilde{\theta} _k(h)$, a microstructure sample of size $\SI{80.2}{\micro\meter} \times \SI{80.2}{\micro\meter} \times \SI{80.2}{\micro\meter}$, corresponding to $401\times 401\times 401$ voxel grid, is generated.
In total a database comprising 495 virtual 3D microstructures is obtained.

\subsection{Effective macroscopic properties and their simulation} \label{subsec: effective macroscopic properties and their estimation}
Note that, analogously to Section~\ref{subsec: morphological descriptors and their estimation}, the effective macroscopic properties, which are defined for 3D images in the following, are computed for model realizations of $\Xi$. In addition, it should be noted that the governing equations used to derive effective macroscopic properties are defined on continuous domains, which we denote by $\Omega \subset \mathbb{R}^3$. However, realizations of $\Xi$ define the transport phases (i.e., the SE and AM phases) on a discrete grid. As the domain $\Omega$ is discretized for numerically computing solutions of the governing equations, this discrepancy is not an issue, since the discretized transport phases in realizations of $\Xi$ can be considered to be the discretized version of the domain $\Omega$. In the following, we therefore introduce the governing equations for the computation of the effective macroscopic properties on the continuous domain. 
\paragraph{Conductivity}
Two key effective properties of ASSB cathodes are the ionic and electronic conductivity, describing the ion and electron transport through the composite microstructure.
Due to the negligible ionic conductivity of the AM and the negligible electronic conductivity of the SE, transport is modeled by considering only one conducting phase at a time~\cite{zhang2018new}.
Under steady-state conditions and assuming electroneutrality as well as the absence of charge accumulation, both electronic and ionic transport can be described within the same mathematical framework.
Although electronic conduction is inherently Ohmic and ionic transport originates from Fick's Law~\cite{fick1855ueber}, the computation of effective ionic and electronic conductivities in microstructure-resolved simulations reduces to solving a steady-state conduction problem in the respective conducting phase.
Therefore, the concept is only introduced for the electronic conductivity.
Let $\Omega\subset \R ^3$ be the continuous domain representing the material phase, where transport occurs.  
Charge transport is described by Ohm's Law
\begin{align}
    J=-\sigma_0 \nabla U \quad\text{in}\; \mathring{\Omega},
\end{align}
where $J\colon \Omega \to \R ^3$ denotes the current density, $\sigma_0>0$ the phase-specific intrinsic conductivity, and $U\colon \Omega \to \R$ the electric potential.
Here, $\mathring{\Omega}$ denotes the interior of $\Omega$.
Conservation of charge requires
\begin{align}
    \nabla \cdot J=0 \quad\text{in}\; \mathring{\Omega},
\end{align}
which yields the equation
\begin{align}
    \nabla \cdot (\sigma_0\nabla U)=0 \quad\text{in}\; \mathring{\Omega}.
\end{align}
Since $\sigma _0$ is constant, this reduces to the Laplace equation
\begin{align}
    \nabla ^2 U=0 \quad\text{in}\; \mathring{\Omega}.
\end{align}
Note that Dirichlet boundary conditions are specified on two opposite external surfaces of $\Omega$, assigning $U$ with the value 0 on one surface and 1 on the opposite surface.
Zero-flux boundary conditions for $U$ are applied to all remaining surfaces, including both the other external faces as well as the internal interfaces between the conducting phase and its complement.
The effective conductivity $\effcond$ is then determined by solving this boundary value problem on the microstructure and relating the respective flux to that of the homogenized problem~\cite{neumann2023data}.
As mentioned above, the same mathematical framework applies for the determination of ionic conductivity.

\paragraph{M-factor}
The M-factor, defined as 
\begin{align}
    M = \frac{\effcond}{\intrcond},
\end{align}
is a useful measure to quantify the influence of microstructure on transport processes.
Here, $\intrcond$ is the intrinsic transport coefficient, representing the ideal transport properties of a uniform bulk material.
In contrast, the effective transport coefficient $\effcond$ accounts for the actual microstructure geometry and incorporates the effects of obstacles, constrictions, and phase connectivity, effectively scaling $\intrcond$ to reflect real transport conditions. 
The M-factor can be applied to both ionic and electronic conductivity, making it a versatile tool for characterizing transport in complex cathode microstructures.
For the phase with volume fraction $\varepsilon$ in which transport occurs, the M-factor satisfies $M\in [0, \varepsilon]$ according to Eq. 21.14 in~\cite{torquato2002random}. 

The effective thermal conductivity and the M-factor with respect to ionic and electronic conductivity are computed by numerical homogenization of the voxelized microstructure using GeoDict~\cite{GeoDict2026}. In GeoDict, the stationary heat equation is solved on a Cartesian grid, and the macroscopic conductivity tensor is obtained from the volume-averaged heat flux under imposed temperature gradients~\cite{WiegmannZemitis2006}.

\subsection{Regression models for structure-property relationships}\label{subsec: regression models}
Pairing geometrical descriptor values computed from a set of 3D images with their corresponding M-factors yields a dataset for regression models, enabling the relation of geometrical descriptors to effective macroscopic properties.
More precisely, for each microstructure we consider the pair $(\varepsilon_{j,p}, S_{j,p}, \mu(\tau_{j,p}), \sigma(\tau_{j,p}), \beta_{j,p})) \times (m_{j,p})$, where $\varepsilon_{j,p},S_{j,p}, \mu(\tau_{j,p}), \sigma(\tau_{j,p}), \beta_{j,p}$ denote the volume fraction, the specific surface area, the mean and standard deviation of the geodesic tortuosity and the constrictivity of phase $p \in \{\text{AM}, \text{SE}\}$ for the j-th ($j = 1, \dots , 495$) microstructure. The M-factor of phase p and the j-th microstructure is denoted by $m_{j,p}$. The dataset of pairs is then split into a training dataset containing 331 microstructures and a test dataset containing 164 microstructures. 

In this work, regression models in the form of analytical formulas are introduced that will be used to establish microstructure-property relationships, i.e., formulas that enable the prediction of the M-factor from geometrical descriptors for the AM and SE. For simplification, the phase under consideration is omitted in the mathematical notation below.
Several regression models have been proposed in the literature to characterize transport-related structure-property relationships.
The simplest approach considers only the volume fraction and is defined as
\begin{align}\label{eq: m1}
    \mfactor_{1,C_1}(V_1) = \varepsilon ^{c_1},
\end{align}
for $C_1=c_1 \in \R_{>0}$ and $V_1 = \varepsilon$. 
Some slight variations of this formula have been investigated in literature~\cite{vadakkepatt2016bruggeman, chung2013validity}. 
A slightly modified version of the more advanced model introduced in \cite{stenzel2016predicting}  additionally incorporates further geometrical descriptors---namely the constrictivity $\beta$ and mean geodesic tortuosity $\mu (\tau)$---is given by
\begin{align} \label{eq: m2}
    \mfactor_{2,C_2}(V_2) = c_1 \varepsilon ^{c_2}\beta^{c_3}\mu (\tau)^{c_4},
\end{align}
with $C_2=(c_1, c_2, c_3, c_4) \in \R^4$ and $V_2=(\varepsilon, \beta, \mu (\tau))$.
Another model, used in~\cite{prifling2021large}, considers both the mean and standard deviation of the geodesic tortuosity as well as the volume fraction and is given by
\begin{align}\label{eq: m3}
    \mfactor_{3,C_3} (V_3) = c_1 \mu(\tau)^{c_2}\sigma (\tau)^{c_3}\varepsilon ^{c_4},
\end{align}
where $C_3=(c_1, c_2, c_3, c_4) \in \R^4$ and $V_3=(\varepsilon, \mu (\tau), \sigma(\tau))$.
Extending $\mfactor_{(3, C_3)}$ by additionally including the constrictivity yields
\begin{align}\label{eq: m4}
    \mfactor_{4,C_4} (V_4) = c_1 \mu(\tau)^{c_2}\sigma (\tau)^{c_3}\varepsilon ^{c_4} \beta ^{c_5},
\end{align}
for $C_4=(c_1, c_2, c_3, c_4, c_5) \in \R^5$ and $V_4=(\varepsilon, \mu (\tau), \sigma(\tau))$.

The parameters of the proposed parametric regression models are estimated by least squares optimization using Scipy's \texttt{minimize} function with the L-BFGS-B algorithm \cite{virtanen2020scipy}.
Specifically, the objective function to be minimized is given by
\begin{align}
    g(C_i) = \sum _{j=1}^{495}\left (m_j-\mfactor_{i,C_i}(V_j) \right )^2,
\end{align}
where $m_j$ denotes the observed values and $\mfactor_{i,C_i}(V_j)$ the corresponding predictions of model $\mfactor _i$, with $i=1,\dots,4,\ j=1,\dots 495$.

In order to evaluate the predictive capabilities of the regression models, the mean absolute percentage error (MAPE) between the predicted and simulated M-factors is considered, as they quantify complementary aspect of model quality---namely the relative prediction error and the explained variance.
For simplicity, the notation $\widehat{m}_{i,j}=\mfactor _{i, C_i} (V_j)$ for $i=1,\dots 4$ and $j=1,\dots, k$ is used in the following.
It is defined as
\begin{align}
    \mape _i = \frac{100}{495} \sum _{j=1}^{495}\left\vert \frac{\widehat{m}_{i,j}-m_{j}}{m_{j}}\right\vert,
\end{align}
where $(m_{1},\dots, m_{k})$ denote the simulated M-factors and $(\widehat{m}_{i,1},\dots, \widehat{m}_{i,k})$ the corresponding predictions computed via $\mfactor _i$, $i=1,\dots , 4$.
Smaller MAPE values indicate a higher predictive accuracy, with a value of zero corresponding to perfect agreement.
Additionally, the coefficient of determination $R^2\in[0,1]$ is evaluated according to
\begin{align}
    R_i ^2 = 1- \frac{\sum\limits _{j=1}^{495}(m_{j}-\widehat{m}_{i,j})^2}{\sum\limits _{j=1}^{495}(m_{j}-\overline{m}_{j})^2}
\end{align}
for $i=1,\dots 4$, where $\bar{y}$ denotes the empirical mean of the ground truth data $m_{1},\dots,m_{495}$.
Values of $R^2$ close to one indicate a strong explanatory power over the model, whereas values close to zero reflect limited predictive capability.
For simplicity, the notation $\mfactor _i=\mfactor_{i,C_i}(V_i)$ for $i=1,\dots, 4$ is used in the following.

\section{Results}
\subsection{Parameter study}\label{sec: param study}
\begin{figure}[H]
    \centering
    \includegraphics[width=0.9\linewidth]{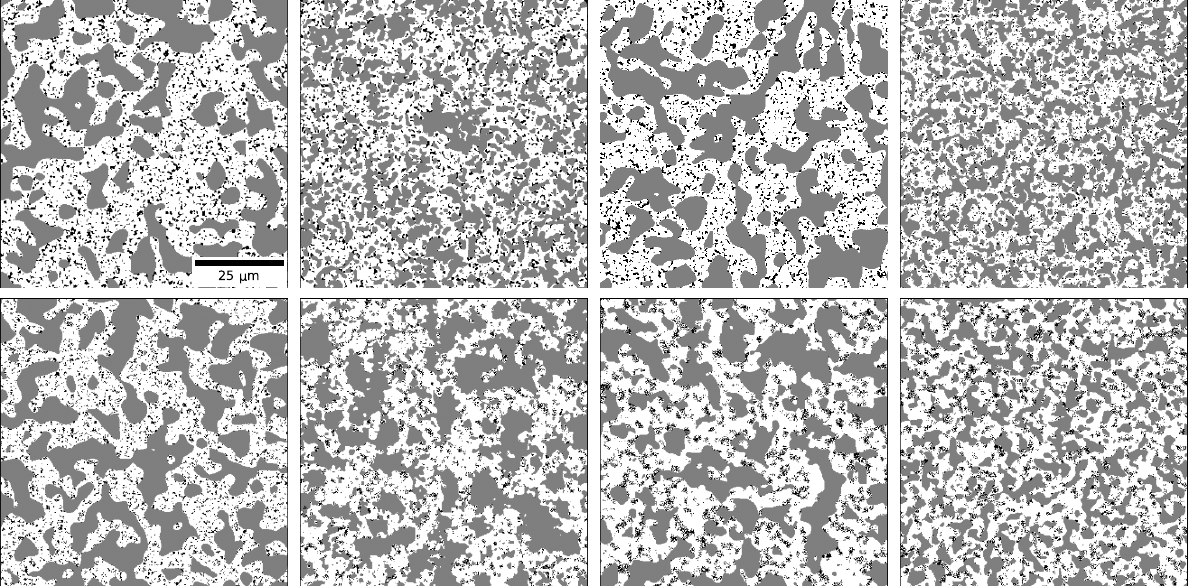}
    \caption{2D cross-sections of exemplary chosen 3D microstructures resulting from the parameter study.}
    \label{fig: 2D slices}
\end{figure}

The results of the parameter study introduced in Section~\ref{subsec: sim study}, employing both interpolation-based and gradient-based approaches, are presented in this section.
Figure~\ref{fig: 2D slices} illustrates representative 2D cross-sections from the generated 3D microstructures.

\begin{figure}[H]
    \centering
    \includegraphics[width=0.85\linewidth]{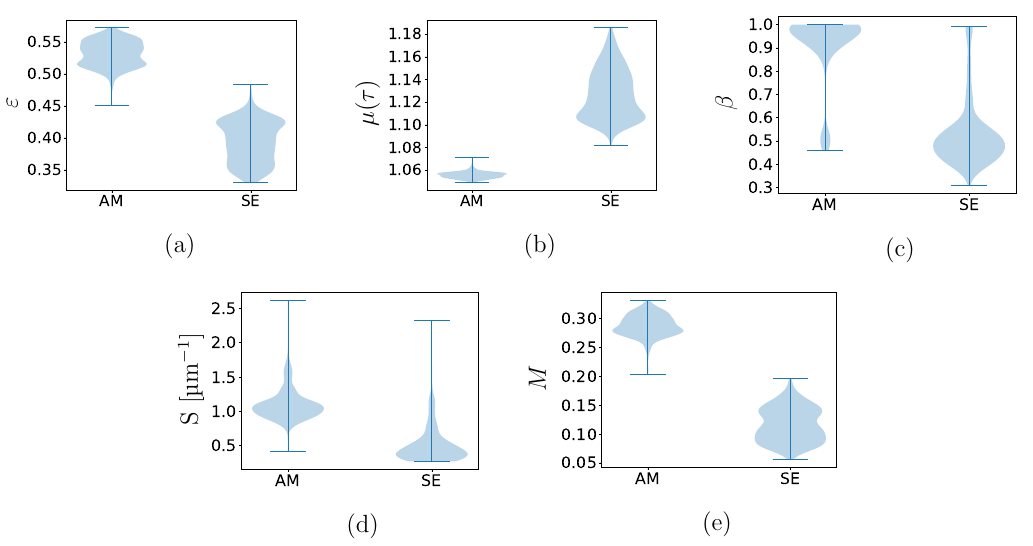}
    \caption{(a) Volume fraction, (b) mean geodesic tortuosity, (c) constrictivity, (d) specific surface area and (e) M-factor associated with the AM and SE of all generated 3D ASSB cathode microstructures of the parameter study.}
    \label{fig: desc spectrum}
\end{figure}

For all $495$ generated 3D ASSB cathode microstructures, several geometrical descriptors are computed, since we later investigate how these these descriptors relate to the M-factor.
In particular, the database of virtual 3D microstructures is designed to cover a wide range of geometrical descriptor values and M-factor values.
Figure~\ref{fig: desc spectrum} presents a quantitative analysis of the volume fraction, mean geodesic tortuosity, constrictivity, and specific surface area, as well as the M-factor, each computed for the AM and SE.

The volume fraction of the AM spans $0.45$ to $0.57$, while that of the SE ranges from $0.33$ to $0.48$.
With respect to the mean geodesic tortuosity, which quantifies the shortest path lengths, only minor variations of about $0.03$ are observed for the AM, whereas $\mu (\tau _\SE)$ exhibits a wider range from $1.081$ to $1.186$.
The constrictivity shows greater variability for both phases, with most values for the AM lying within $[0.85,1]$, while those for the SE range from $0.31$ to $0.99$.
A similar trend can be seen in the specific surface area, where $S_\AM$ ranges from $\SI{0.42}{\micro\meter}$ to $\SI{2.61}{\micro\meter}$ and is predominantly clustered in the range $[\SI{0.7}{\micro\meter}, \SI{1.7}{\micro\meter}]$. 
For the SE, $S_\SE$ ranges from $\SI{0.27}{\micro\meter}^{-1}$ to $ \SI{2.33}{\micro\meter}^{-1}$, with the majority of values lying in $[\SI{0.27}{\micro\meter}^{-1}, \SI{1.35}{\micro\meter}^{-1}]$.
Finally, the M-factor for the AM spans from $0.20$ to $0.33$ and that for the SE from $0.06$ to $0.20$, indicating that the differences observed in the geometrical descriptors discussed above also influence the M-factor.
For further details, see Section~\ref{sec: discussion}.

\subsection{Structure-property relationships} \label{sec: struc-prop rel}
\begin{figure}[H]
    \centering
    \includegraphics[width=0.7\linewidth]{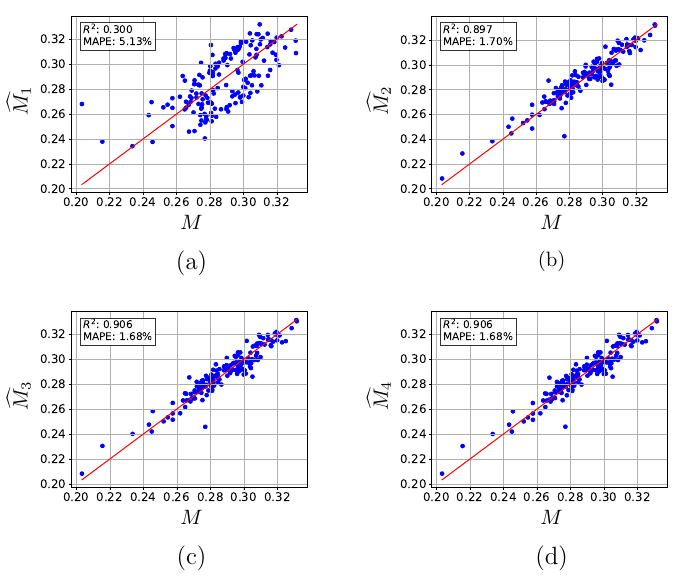}
    \caption{Predicted M-factors over numerically simulated $M$-factors with respect to the AM for all 164 virtual ASSB cathode microstructures of the test dataset using the prediction formulas $\mfactor _1$ (a), $\mfactor _2$ (b), $\mfactor _3$ (c) and $\mfactor _4$ (d). Red line ($\hat{M} = M$) serves as reference.}
    \label{fig: regressions AM}
\end{figure}

The virtual 3D microstructures presented above are divided into a training dataset comprising 331 microstructures and a test dataset of 164 microstructures.
The training dataset is used to calibrate the regression models introduced in Section~\ref{subsec: regression models}.
The resulting coefficients for these models are presented in Table~\ref{tab: coeffs regression} for both the AM and the SE.
\begin{table}[H]
    \centering
    \caption{Coefficients of the regression models $\mfactor _1$, $\mfactor _2$, $\mfactor _3$ and $\mfactor _4$ for the AM and SE obtained by least-squares optimization using the training dataset of 331 microstructures. Additionally, the corresponding $R^2$ values computed for the test dataset of 164 microstructures are presented.}
    \vspace{0.2cm}
    \begin{tabular}{c|llll}
         & $\mfactor _1$ & $\mfactor _2$ & $\mfactor _3$ & $\mfactor _4$ \\
         \hline
         \multirow{5}{*}{AM} & $c_1=1.98$  & $c_1 = 0.99$ & $c_1=1.10$ & $c_1=1.10$\\
         & & $c_2=0.11$ & $c_2=-19.60$ & $c_2=-19.60$\\
         & & $c_3=-0.002$ & $c_3=0.03$ & $c_3=0.03$\\
         & & $c_4=-21.52$ & $c_4=0.24$ & $c_4=0.24$ \\
         & & & & $c_5=-0.002$\\[0.2cm]
         & $R^2=0.300$ & $R^2=0.897$ & $R^2=0.906$ & $R^2=0.906$\\
         \hline
         \multirow{5}{*}{SE} & $c_1=2.32$  & $c_1 = 2.74$ & $c_1=3.99$ & $c_1=4.36$\\
         & & $c_2=3.88$ & $c_2=-0.54$ & $c_2=-1.42$\\
         & & $c_3=-0.09$ & $c_3=0.13$ & $c_3=0.16$\\
         & & $c_4=3.09$ & $c_4=3.26$ & $c_4=3.08$ \\
         & & & & $c_5=0.06$\\[0.2cm]
         & $R^2=0.843$ & $R^2=0.882$ & $R^2=0.912$ & $R^2=0.911$\\
    \end{tabular}
    
    \label{tab: coeffs regression}
\end{table}
The results of all four regression models for the M-factor computed for the AM are visualized in Figure~\ref{fig: regressions AM}, where the predicted M-factors are plotted against those obtained from numerical simulations, along with the corresponding coefficient of determination $R^2$ and their mean absolute percentage errors.
The first model, $\mfactor _1$ given in Eq.~\eqref{eq: m1}, which only considers the volume fraction, yields a comparatively low $R^2$ of $0.300$ and MAPE of approximately $5\%$, as shown in Figure~\ref{fig: regressions AM}a).
Substantial improvement of both error measures is observed for $\mfactor _2$, given in Eq.~\eqref{eq: m2}, see Figure~\ref{fig: regressions AM}b).
In particular, a value of $R^2 = 0.897$ is achieved for the coefficient of determination and $\mape = 1.70\%$.
Similar results are obtained with regression model $\mfactor _3$, given in Eq.~\eqref{eq: m3}, where additionally to the volume fraction $\varepsilon$, the mean and standard deviation of the geodesic tortuosity are included, yielding $R^2=0.906$ and $\mape = 1.68\%$, see Figure~\ref{fig: regressions AM}c).
Figure~\ref{fig: regressions AM}d) shows that the same values are achieved for model $\mfactor _4$, stated in Eq.~\eqref{eq: m4}, that combines all geometrical descriptors used for $\mfactor _2$ and $\mfactor _3$.

\begin{figure}[H]
    \centering
    \includegraphics[width=0.7\linewidth]{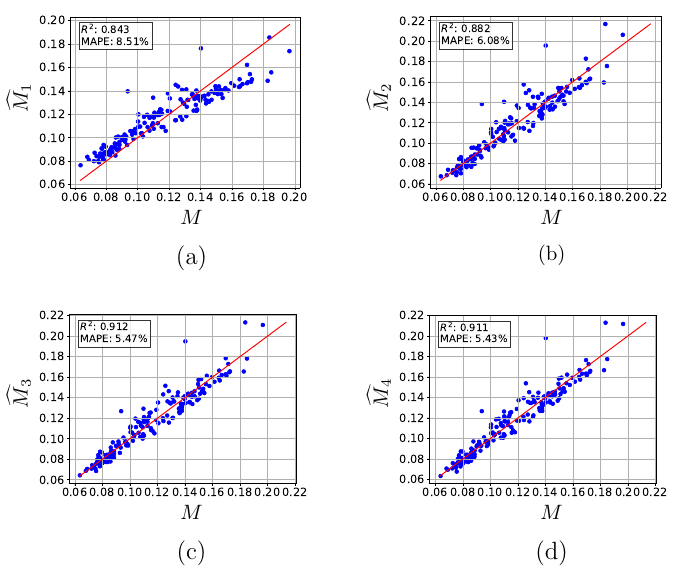}
    \caption{Predicted M-factors over numerically simulated M-factors for the SE for all 164 virtual ASSB cathode microstructures of the test dataset using the prediction formulas $\mfactor _1, \mfactor _2,\mfactor _3$ and $\mfactor _4$.}
    \label{fig: regressions SE}
\end{figure}

The results of the fitted regression models with respect to the SE are shown in Figure~\ref{fig: regressions SE}.
The first model $\mfactor _1$, given in Eq.~\eqref{eq: m1}, which attempts to find a direct connection between the volume fraction $\varepsilon _\SE$ and the M-factor, achieves an $R^2=0.843$ and a MAPE of $8.51\%$, see Figure~\ref{fig: regressions SE}a), which is significantly better than the corresponding result for the AM.
The second model, $\mfactor _2$, which additionally incorporates the constrictivity and mean geodesic tortuosity, yields a $R^2=0.882$ and a MAPE of $6.08\%$, as can be seen in Figure~\ref{fig: regressions SE}b).
Models $\mfactor _3$ and $\mfactor _4$ further increase the coefficient of determination to $R^2=0.912$ and $0.911$, respectively, while the MAPE decreases to $5.47\%$ and $5.43\%$, see Figure~\ref{fig: regressions SE}c) and d).

\section{Discussion}\label{sec: discussion}
In this section, the methods and results presented in this paper are discussed. 
Specifically, the selected methods used to generate a large database of virtual yet realistic ASSB cathode microstructures are evaluated, followed by a discussion of the analysis with respect to several geometrical descriptors. 
Finally, the results of the fitted regression models are examined in order to quantify and interpret the underlying structure-property relationships.

Typically, when aiming to generate a large database of virtual image data based on a parametric stochastic geometry model---particularly for investigating structure-property relationships---the input parameters are systematically varied to cover a broad spectrum of differently structured microstructures.
In many cases, the volume fraction can be controlled directly, making it possible to generate samples that are uniformly distributed with respect to this fundamental geometrical descriptor as demonstrated in~\cite{prifling2021large}. 
However, other descriptors---especially those closely related to transport properties, such as geodesic tortuosity and constrictivity---often cannot be controlled explicitly, a limitation that becomes more pronounced for complex stochastic geometry models such as the one employed in this work.
Consequently, it can be difficult to generate microstructures that both remain close to experimentally measured ones and allow for a controlled variation of specific geometrical descriptors.
Therefore, the gradient-based approach presented in this paper offers a new perspective on systematic parameter variation and generation of large databases of virtual 3D microstructures.

The gradient-based approach is illustrated in Figure~\ref{fig: gradient approach schmee} for the specific surface area of the second interpolated parameter vector between BM03 and BM10.
For this exemplarily chosen reference parameter vector, the specific surface area of a corresponding model realization is $\SI{0.30}{\micro\meter}^{-1}$.
For $h=-0.1$, a value of $\SI{0.81}{\micro\meter}^{-1}$ is obtained, and for $h=-0.2$ $\SI{1.29}{\micro\meter}^{-1}$.
In the opposite direction of the gradient, the specific surface area remains nearly unchanged, with values of $\SI{0.28}{\micro\meter}^{-1}$ for both $h=0.1$ and $h=0.2$.
As shown in Figure~\ref{fig: desc spectrum}, this approach enables the generation of a database comprising a variety of distinct morphologies.
The volume fraction of the AM ranges from $0.45$ to $0.57$, while that of the SE ranges from $0.33$ to $0.48$.
The mean geodesic tortuosity of the SE can reach values from $1.08$ to $1.18$ can be achieved. 
In contrast, the mean geodesic tortuosity of the AM is comparatively low, which is expected, as the AM phase is highly connected due to the dense and tightly packed AM particles, implying that the shortest paths through the AM are almost straight~\cite{park2022geometrical}.
Higher AM tortuosity would be undesirable, as it would correspond to a highly porous and poorly connected AM network.
With the same reasoning, it can be explained that most of the AM constrictivity values are clustered between $0.9$ and $1$.
In comparison, the SE constrictivity, as well as the specific surface area of both the AM and SE, exhibit wider value ranges, reflecting greater variability in the geometry.
The M-factor exhibits a good spread for both the AM and the SE, providing a suitable basis for the regression models.
The noticeable difference in the range of M-factor values between the AM and the SE can be explained by the fact that $M$ is bounded above by the corresponding volume fraction.
Although the ranges overlap, the volume fraction of the AM extends to higher values than that of the SE, which in turn allows the M-factor values for the AM to reach higher values.

The evaluation of M-factors both through different prediction models and numerical simulations shows that the volume fraction alone is insufficient to predict the M-factor.
As shown in Figure~\ref{fig: regressions AM}a), this results in a low coefficient of determination, $R^2=0.300$ for the AM, highlighting that other geometrical descriptors beyond the volume fraction significantly influence transport.
In contrast, for the SE, a higher $R^2=0.843$ is achieved, indicating that the volume fraction is the dominant factor controlling ionic conductivity in this phase.
This observation is supported by the fact that the coefficient $c_1$ in $\mfactor _1$ is higher for the SE, specifically $c_1=2.32$, than for the AM, where we have $c_1=1.98$, see Table~\ref{tab: coeffs regression}.
Incorporating the constrictivity and the mean geodesic tortuosity, as done for $\mfactor _2$, yields significantly improved results---an effect that is far more pronounced for the AM than for the SE.
This is expected, as both descriptors directly impact transport pathways, particularly in the more complex microstructure of the AM.
The fact that the coefficient of determination $R^2$ only increases slightly---by $0.009$ for the AM and by $0.03$ for the SE---suggests that including either the standard deviation of the geodesic tortuosity or the constrictivity alongside the volume fraction and the mean geodesic tortuosity does not make a substantial difference.
The models $\mfactor _3$ and $\mfactor _4$ show similar results for the AM, with $R^2$ around $0.906$ and $\mape =1.68\%$, indicating that the constrictivity provides limited additional information for predicting $M$.
For the M-factors of the SE, a small decrease in the $R^2$ value of $0.001$ is observed, which may indicate that $\mfactor _4$ is slightly overfitting with respect to the training data.
These findings imply that, for accurate prediction of M-factors, incorporating mean geodesic tortuosity is essential for the AM, while for the SE, the volume fraction alone captures most of the variability.

Some of these M-factor prediction models have already been investigated in the literature.
Comparing the coefficients they computed to those obtained in this work shows that there are some similarities.
For model $\mfactor _3$, when considering the AM, $c_3$ is nearly identical to the value reported in~\cite{prifling2021large} and only a minor difference can be observed in $c_1$, namely $c_1=1.18$ in~\cite{prifling2021large} and $c_1 = 1.10$ in our case.

When looking at the coefficients in Table~\ref{tab: coeffs regression}, it is observable that the constrictivity only has a minor influence on the M-factor of the AM as the corresponding coefficients are very small, precisely $c_3=-0.002$ for $\mfactor _2$ and $c_5=-0.002$ for $\mfactor _4$.
For SE, on the other hand, the coefficients are slightly higher with values of $c_3=-0.09$ and $c_5=-0.06$ indicating that the constrictivity is more important for the M-factor of the  SE than it is for that of the AM.
A similar phenomenon can be seen for the standard deviation of the tortuosity $\sigma (\tau)$. 
It seems to have only a minor influence on the M-factor of the AM, with coefficients of $0.03$ for $\mfactor _3$ and $\mfactor _4$, while for the SE they amount to $c_3=0.13$ for $\mfactor _3$ and $c_3=0.16$ for $\mfactor _4$.

The best trade-off between model simplicity and predictive accuracy of the M-factors is achieved by $\mfactor _3$ for both the AM and SE, which incorporates the volume fraction as well as the mean and standard deviation of the geodesic tortuosity.
More precisely, for the AM, the corresponding expression is given by 
\begin{align}\label{eq: reg opt AM}
    \mfactor _3 = 1.10 \mu(\tau _\AM) ^{-19.60}\sigma (\tau _\AM)^{0.03}\varepsilon_\AM ^{0.24}
\end{align}
and for the SE by 
\begin{align}\label{eq: reg opt SE}
    \mfactor _3 = 3.99 \mu(\tau _\SE) ^{-0.54}\sigma (\tau _\SE)^{0.13}\varepsilon_\SE ^{3.26}.
\end{align}

\section{Conclusions}
In the present paper, an approach is introduced that enables systematic parameter variation and thus the generation of a large database of virtual 3D ASSB cathode microstructures that have not yet been observed experimentally.
For these microstructures, effective macroscopic properties such as ionic and electronic conductivity can be computed by means of numerical simulations.
This framework enables the investigation of structure-property relationships by identifying correlations between these effective transport properties and geometrical descriptors, while avoiding the high costs and time consuming procedures associated with the production and imaging of experimental samples.

To implement this virtual materials testing framework, a stochastic microstructure model is employed.
Specifically, a simplified version of the model introduced in~\cite{furat2025generative}, which is based on excursion sets of random fields and calibrated using a generative adversarial framework, is used.

Usually, when generating a large database of virtual microstructures using a stochastic model, the volume fraction can be directly influenced, which makes it possible to generate samples that are uniformly distributed with respect to this geometrical descriptor, as, for example, in~\cite{prifling2021large}.
However, typically it is difficult to specify parameters of stochastic models such that generated microstructures exhibit predefined values for other descriptors---especially those closely linked to transport processes, for example geodesic tortuosity and constrictivity.
This applies in particular to more complex stochastic models, as is the case with the model used in this paper.
Combining interpolation between fitted parameter vectors corresponding to three experimentally measured images with a gradient-based approach allows this limitation to be overcome. 
Based on each interpolated point, the difference quotient with respect to a specific geometrical descriptor can be evaluated for each entry of a parameter vector $\theta \in \Theta$ and thus leads to an approximation of the gradient.
This enables a systematic variation of  cathode microstructures with respect to those geometrical descriptors.

Computing ionic and electronic conductivity for all generated structures using GeoDict enables the investigation of structure-property relationships.
More specifically, several regression models---based on different combinations of geometrical descriptors---were calibrated to the training dataset to quantify the influence of volume fraction, geodesic tortuosity and constrictivity on effective transport properties of these cathode microstructures.
Based on the fitted regression model coefficients, the mean absolute percentage error, and the coefficient of determination $R^2$ are computed for the test dataset.
The results show that the volume fraction has a substantially greater influence on the M-factor of the SE than on that of the AM.
Furthermore, the mean geodesic tortuosity also proves to be an important geometrical descriptor when examining its effect on ionic and electronic conductivity.
In contrast, the constrictivity and the standard deviation of the mean geodesic tortuosity appear to have a smaller impact on these effective transport properties.
Nevertheless, at least one of these two geometrical descriptors, together with the volume fraction and the mean geodesic tortuosity, is required to obtain well-fitting regression models.
The best trade-off between model complexity and predictive accuracy of the M-factor is given by $\mfactor _3$ for both the AM and SE comprising the volume fraction as well as the mean and standard deviation of the Geodesic tortuosity, see Eq.~\eqref{eq: reg opt AM} and~\eqref{eq: reg opt SE}.

These structure-property relationships build the foundation for future inverse microstructure design, which aims to identify the optimal values of relevant geometrical descriptors required to achieve a desired set of effective material properties, such as maximized ionic conductivity, sufficient electronic conductivity, and minimized tortuosity.
Instead of predicting properties from a given microstructure, the structure-property relationship is inverted by formulating an optimization problem.
The goal is to identify the optimal geometrical descriptor ranges and model parameters that produce such desired macroscopic transport properties.

\section*{Acknowledgement}
This research is funded by the German Federal Ministry of Research, Technology and Space (BMFTR) under grant number 03XP0562.

% \bibliographystyle{elsarticle-num}
% \bibliography{bibliography}

\end{document}

%% file: macros.tex
\definecolor{ulmgruen}{rgb}{0.3272,0.6667,0.1098}

\newcommand{\R}{\mathbb{R}}
\newcommand{\Z}{\mathbb{Z}}

\newcommand{\AM}{\mathrm{AM}}
\newcommand{\SE}{\mathrm{SE}}
\newcommand{\PS}{\mathrm{P}}

\newcommand{\sfaAM}{f_{S_{\mathrm{AM}}}}
\DeclareSIUnit{\weightpercent}{\text{w\%}}

\newcommand{\excOne}{\Xi (t,1)}
\newcommand{\excTwo}{\Xi (t,2)}
\newcommand{\excThree}{\Xi (t,3)}

\newcommand{\intrcond}{\sigma _0}
\newcommand{\effcond}{\sigma _{\mathrm{eff}}}
\newcommand{\mfactor}{\widehat{M}}
\newcommand{\mape}{\mathrm{MAPE}}

\newcommand{\aniso}{a}

%% file: main.bbl
\begin{thebibliography}{10}
\expandafter\ifx\csname url\endcsname\relax
  \def\url#1{\texttt{#1}}\fi
\expandafter\ifx\csname urlprefix\endcsname\relax\def\urlprefix{URL }\fi
\expandafter\ifx\csname href\endcsname\relax
  \def\href#1#2{#2} \def\path#1{#1}\fi

\bibitem{korthauer2018lithium}
R.~Korthauer, Lithium-Ion Batteries: Basics and Applications, Springer, 2018.

\bibitem{passerini2020batteries}
S.~Passerini, D.~Bresser, A.~Moretti, A.~Varzi, Batteries: Present and Future Energy Storage Challenges, J. Wiley \& Sons, 2020.

\bibitem{aziam2022solid}
H.~Aziam, B.~Larhrib, C.~Hakim, N.~Sabi, H.~B. Youcef, I.~Saadoune, Solid-state electrolytes for beyond lithium-ion batteries: A review, Renewable and Sustainable Energy Reviews 167 (2022) 112694.

\bibitem{chen2021research}
J.~Chen, J.~Wu, X.~Wang, A.~Zhou, Z.~Yang, Research progress and application prospect of solid-state electrolytes in commercial lithium-ion power batteries, Energy Storage Materials 35 (2021) 70--87.

\bibitem{lim2020review}
H.-D. Lim, J.-H. Park, H.-J. Shin, J.~Jeong, J.~T. Kim, K.-W. Nam, H.-G. Jung, K.~Y. Chung, A review of challenges and issues concerning interfaces for all-solid-state batteries, Energy Storage Materials 25 (2020) 224--250.

\bibitem{janek2023challenges}
J.~Janek, W.~G. Zeier, Challenges in speeding up solid-state battery development, Nature Energy 8 (2023) 230--240.

\bibitem{ren2023oxide}
Y.~Ren, T.~Danner, A.~Moy, M.~Finsterbusch, T.~Hamann, J.~Dippell, T.~Fuchs, M.~M{\"u}ller, R.~Hoft, A.~Weber, , L.~A. Curtiss, P.~Zapol, M.~Klenk, A.~T. Ngo, P.~Barai, B.~C. Wood, R.~Shi, L.~F. Wan, T.~W. Heo, M.~Engels, J.~Nanda, F.~H. Richter, A.~Latz, V.~Srinivasan, J.~Janek, J.~Sakamoto, E.~D. Wachsman, D.~Fattakhova-Rohlfing, Oxide-based solid-state batteries: a perspective on composite cathode architecture, Advanced Energy Materials 13 (2023) 2201939.

\bibitem{zhang2018synthesis}
D.~Zhang, X.~Cao, D.~Xu, N.~Wang, C.~Yu, W.~Hu, X.~Yan, J.~Mi, B.~Wen, L.~Wang, L.~Zhang, Synthesis of cubic \ce{Na_3SbS_4} solid electrolyte with enhanced ion transport for all-solid-state sodium-ion batteries, Electrochimica Acta 259 (2018) 100--109.

\bibitem{asheri2024microstructure}
A.~Asheri, S.~Rezaei, V.~Glavas, B.-X. Xu, Microstructure impact on chemo-mechanical fracture of polycrystalline lithium-ion battery cathode materials, Engineering Fracture Mechanics 309 (2024) 110370.

\bibitem{vu2023towards}
T.-S. Vu, M.-Q. Ha, D.-N. Nguyen, V.-C. Nguyen, Y.~Abe, T.~Tran, H.~Tran, H.~Kino, T.~Miyake, K.~Tsuda, H.-C. Dam, Towards understanding structure--property relations in materials with interpretable deep learning, npj Computational Materials 9 (2023) 215.

\bibitem{marmet2024multiscale}
P.~Marmet, L.~Holzer, T.~Hocker, H.~Bausinger, J.~G. Grolig, A.~Mai, J.~M. Brader, G.~K. Boiger, Multiscale-multiphysics model for optimization of novel ceramic {MIEC} solid oxide fuel cell electrodes, The International Journal of Multiphysics 18 (2024) 58--83.

\bibitem{clausnitzer2023optimizing}
M.~Clausnitzer, R.~M{\"u}cke, F.~Al-Jaljouli, S.~Hein, M.~Finsterbusch, T.~Danner, D.~Fattakhova-Rohlfing, O.~Guillon, A.~Latz, Optimizing the composite cathode microstructure in all-solid-state batteries by structure-resolved simulations, Batteries \& Supercaps 6 (2023) e202300167.

\bibitem{park2020digital}
J.~Park, K.~T. Kim, D.~Y. Oh, D.~Jin, D.~Kim, Y.~S. Jung, Y.~M. Lee, Digital twin-driven all-solid-state battery: unraveling the physical and electrochemical behaviors, Advanced Energy Materials 10 (2020) 2001563.

\bibitem{liu2025fib}
Z.~Liu, S.~Bai, S.~Burke, J.~N. Burrow, R.~Geurts, C.-J. Huang, C.~Jiao, H.-B. Lee, Y.~S. Meng, L.~Nov{\'a}k, B.~Winiarski, J.~Wang, K.~Wu, M.~Zhang, {FIB-SEM}: emerging multimodal/multiscale characterization techniques for advanced battery development, Chemical Reviews 125 (2025) 5228--5281.

\bibitem{carazo2015three}
J.~Carazo, C.~Sorzano, J.~Ot{\'o}n, R.~Marabini, J.~Vargas, Three-dimensional reconstruction methods in single particle analysis from transmission electron microscopy data, Archives of Biochemistry and Biophysics 581 (2015) 39--48.

\bibitem{maire2014quantitative}
E.~Maire, P.~J. Withers, Quantitative {X}-ray tomography, International Materials Reviews 59 (2014) 1--43.

\bibitem{ohser2000statistical}
J.~Ohser, F.~M{\"u}cklich, Statistical Analysis of Microstructures in Materials Science, J. Wiley \& Sons, 2000.

\bibitem{stenzel2017big}
O.~Stenzel, O.~Pecho, L.~Holzer, M.~Neumann, V.~Schmidt, Big data for microstructure-property relationships: A case study of predicting effective conductivities, AIChE Journal 63 (2017) 4224--4232.

\bibitem{prifling2021large}
B.~Prifling, M.~R{\"o}ding, P.~Townsend, M.~Neumann, V.~Schmidt, Large-scale statistical learning for mass transport prediction in porous materials using 90,000 artificially generated microstructures, Frontiers in Materials 8 (2021) 786502.

\bibitem{neumann2020quantifying}
M.~Neumann, O.~Stenzel, F.~Willot, L.~Holzer, V.~Schmidt, Quantifying the influence of microstructure on effective conductivity and permeability: Virtual materials testing, International Journal of Solids and Structures 184 (2020) 211--220.

\bibitem{nam2025spotlighting}
S.~W. Nam, D.~Lee, E.~Choi, J.~Yeom, S.~H. Choi, D.-J. Yoo, Spotlighting composite cathode heterogeneity: challenges and strategies for all-solid-state batteries, ACS Applied Energy Materials 8 (2025) 6876--6888.

\bibitem{kim2023synergistic}
J.~S. Kim, S.~Jung, H.~Kwak, Y.~Han, S.~Kim, J.~Lim, Y.~M. Lee, Y.~S. Jung, Synergistic halide-sulfide hybrid solid electrolytes for ni-rich cathodes design guided by digital twin for all-solid-state li batteries, Energy Storage Materials 55 (2023) 193--204.

\bibitem{kim2020diffusion}
J.~Y. Kim, J.~Park, M.~J. Lee, S.~H. Kang, D.~O. Shin, J.~Oh, J.~Kim, K.~M. Kim, Y.-G. Lee, Y.~M. Lee, Diffusion-dependent graphite electrode for all-solid-state batteries with extremely high energy density, ACS Energy Letters 5~(9) (2020) 2995--3004.

\bibitem{virtualMaterialTesting}
G.~Gaiselmann, M.~Neumann, V.~Schmidt, O.~Pecho, T.~Hocker, L.~Holzer, Quantitative relationships between microstructure and effective transport properties based on virtual materials testing, AIChE Journal 60 (2014) 1983--1999.

\bibitem{jeziorski2024stochastic}
N.~Jeziorski, C.~Redenbach, Stochastic geometry models for texture synthesis of machined metallic surfaces: sandblasting and milling, Journal of Mathematics in Industry 14 (2024) 17.

\bibitem{theodon2024vox}
L.~Th{\'e}odon, J.~Debayle, C.~Coufort-Saudejaud, Vox-storm: A stochastic 3{D} model based on a dual voxel-mesh architecture for the morphological characterization of aggregates, Powder Technology 444 (2024) 119983.

\bibitem{furat2021artificial}
O.~Furat, L.~Petrich, D.~P. Finegan, D.~Diercks, F.~Usseglio-Viretta, K.~Smith, V.~Schmidt, Artificial generation of representative single li-ion electrode particle architectures from microscopy data, npj Computational Materials 7 (2021) 105.

\bibitem{neumann2023data}
M.~Neumann, S.~E. Wetterauer, M.~Osenberg, A.~Hilger, P.~Gr{\"a}fensteiner, A.~Wagner, N.~Bohn, J.~R. Binder, I.~Manke, T.~Carraro, V.~Schmidt, A data-driven modeling approach to quantify morphology effects on transport properties in nanostructured {NMC} particles, International Journal of Solids and Structures 280 (2023) 112394.

\bibitem{neumann2019pluri}
M.~Neumann, M.~Osenberg, A.~Hilger, D.~Franzen, T.~Turek, I.~Manke, V.~Schmidt, On a pluri-gaussian model for three-phase microstructures, with applications to 3{D} image data of gas-diffusion electrodes, Computational Materials Science 156 (2019) 325--331.

\bibitem{marmet2023stochastic}
P.~Marmet, L.~Holzer, T.~Hocker, V.~Muser, G.~K. Boiger, M.~Fingerle, S.~Reeb, D.~Michel, J.~M. Brader, Stochastic microstructure modeling of {SOC} electrodes based on a pluri-gaussian method, Energy Advances 2 (2023) 1942--1967.

\bibitem{inoue2017numerical}
G.~Inoue, M.~Kawase, Numerical and experimental evaluation of the relationship between porous electrode structure and effective conductivity of ions and electrons in lithium-ion batteries, Journal of Power Sources 342 (2017) 476--488.

\bibitem{fohst2022influence}
S.~F{\"o}hst, S.~Osterroth, F.~Arnold, C.~Redenbach, Influence of geometry modifications on the permeability of open-cell foams, AIChE Journal 68 (2022) e17446.

\bibitem{daubner2024microstructure}
S.~Daubner, B.~Nestler, Microstructure characterization of battery materials based on voxelated image data: Computation of active surface area and tortuosity, Journal of The Electrochemical Society 171 (2024) 120514.

\bibitem{prifling2019parametric}
B.~Prifling, D.~Westhoff, D.~Schmidt, H.~Markoetter, I.~Manke, V.~Knoblauch, V.~Schmidt, Parametric microstructure modeling of compressed cathode materials for li-ion batteries, Computational Materials Science 169 (2019) 109083.

\bibitem{stenzel2016predicting}
O.~Stenzel, O.~Pecho, L.~Holzer, M.~Neumann, V.~Schmidt, Predicting effective conductivities based on geometric microstructure characteristics, AIChE Journal 62 (2016) 1834--1843.

\bibitem{goodfellow2014generative}
I.~J. Goodfellow, J.~Pouget-Abadie, M.~Mirza, B.~Xu, D.~Warde-Farley, S.~Ozair, A.~Courville, Y.~Bengio, Generative adversarial nets, Advances in neural information processing systems 27 (2014).

\bibitem{furat2025generative}
O.~Furat, S.~Weber, A.~Dufter, J.~Schubert, R.~Rekers, M.~Luczak, E.~Glatt, A.~Wiegmann, J.~Janek, A.~Bielefeld, V.~Schmidt, Generative adversarial framework to calibrate excursion set models for the 3{D} morphology of all-solid-state battery cathodes, Advanced Intelligent Systems (2025) 2500572.

\bibitem{anjaData}
P.~Minnmann, J.~Schubert, S.~Kremer, R.~Rekers, S.~Burkhardt, R.~Ruess, A.~Bielefeld, F.~H. Richter, J.~Janek, Editors’ choice—visualizing the impact of the composite cathode microstructure and porosity on solid-state battery performance, Journal of The Electrochemical Society 171 (2024) 060514.

\bibitem{zou2025inverse}
T.~Zou, J.~Shi, M.~Wang, C.~Lin, Inverse design of structured electrodes in lithium metal batteries: Integrated high-throughput phase-field modeling and machine learning, Advanced Functional Materials (2025) e12788.

\bibitem{StGeoAppl}
S.~N. Chiu, D.~Stoyan, W.~S. Kendal, J.~Mecke, Stochastic Geometry and Its Applications, J. Wiley \& Sons, 2013.

\bibitem{sfa_schladitz2006}
K.~Schladitz, J.~Ohser, W.~Nagel, Measuring intrinsic volumes in digital 3{D} images, in: Discrete Geometry for Computer Imagery: 13th International Conference, DGCI 2006, Szeged, Hungary, October 25-27, 2006. Proceedings 13, Springer, 2006, pp. 247--258.

\bibitem{TortConstr}
M.~Neumann, C.~Hirsch, J.~Stan{\v{e}}k, V.~Bene{\v{s}}, V.~Schmidt, Estimation of geodesic tortuosity and constrictivity in stationary random closed sets, Scandinavian Journal of Statistics 46 (2019) 848--884.

\bibitem{dijkstra_jungnickel2005graphs}
D.~Jungnickel, Graphs, Networks and Algorithms, Springer, 2005.

\bibitem{soille1999morphological}
P.~Soille, Morphological Image Analysis: Principles and Applications, Vol.~2, Springer, 1999.

\bibitem{munch2008contradicting}
B.~M{\"u}nch, L.~Holzer, Contradicting geometrical concepts in pore size analysis attained with electron microscopy and mercury intrusion, Journal of the American Ceramic Society 91 (2008) 4059--4067.

\bibitem{zhang2018new}
Z.~Zhang, Y.~Shao, B.~Lotsch, Y.-S. Hu, H.~Li, J.~Janek, L.~F. Nazar, C.-W. Nan, J.~Maier, M.~Armand, L.~Chen, New horizons for inorganic solid state ion conductors, Energy \& Environmental Science 11 (2018) 1945--1976.

\bibitem{fick1855ueber}
A.~Fick, Ueber diffusion, Annalen der Physik 170 (1855) 59--86.

\bibitem{torquato2002random}
S.~Torquato, Random Heterogeneous Materials: Microstructure and Macroscopic Properties, Vol.~16, Springer, 2002.

\bibitem{GeoDict2026}
{Math2Market GmbH}, \href{https://doi.org/10.30423/release.geodict2026}{{GeoDict Simulation Software, Release 2026}} (2026).
\newblock \href {https://doi.org/10.30423/release.geodict2026} {\path{doi:10.30423/release.geodict2026}}.
\newline\urlprefix\url{https://doi.org/10.30423/release.geodict2026}

\bibitem{WiegmannZemitis2006}
A.~Wiegmann, A.~Zemitis, {EJ-HEAT: A Fast Explicit Jump Harmonic Averaging Solver for the Effective Heat Conductivity of Composite Materials}, Fraunhofer ITWM Report~94, Fraunhofer Institute for Industrial Mathematics (ITWM), Kaiserslautern, Germany (2006).

\bibitem{vadakkepatt2016bruggeman}
A.~Vadakkepatt, B.~Trembacki, S.~R. Mathur, J.~Y. Murthy, Bruggeman's exponents for effective thermal conductivity of lithium-ion battery electrodes, Journal of the Electrochemical Society 163 (2016) A119--A130.

\bibitem{chung2013validity}
D.-W. Chung, M.~Ebner, D.~R. Ely, V.~Wood, R.~Edwin~Garc{\'\i}a, Validity of the bruggeman relation for porous electrodes, Modelling and Simulation in Materials Science and Engineering 21 (2013) 074009.

\bibitem{virtanen2020scipy}
P.~Virtanen, R.~Gommers, T.~E. Oliphant, M.~Haberland, T.~Reddy, D.~Cournapeau, E.~Burovski, P.~Peterson, W.~Weckesser, J.~Bright, S.~J. van~der Walt, M.~Brett, J.~Wilson, K.~J. Millman, N.~Mayorov, A.~R.~J. Nelson, E.~Jones, R.~Kern, E.~Larson, C.~J. Carey, {\.I}.~Polat, Y.~Feng, E.~W. Moore, J.~VanderPlas, D.~Laxalde, J.~Perktold, R.~Cimrman, I.~Henriksen, E.~A. Quintero, C.~R. Harris, A.~M. Archibald, A.~H. Ribeiro, F.~Pedregosa, P.~van Mulbregt, Scipy 1.0: fundamental algorithms for scientific computing in python, Nature Methods 17 (2020) 261--272.

\bibitem{park2022geometrical}
S.-Y. Park, J.~Jeong, H.-C. Shin, Geometrical effect of active material on electrode tortuosity in all-solid-state lithium battery, Applied Sciences 12 (2022) 12692.

\end{thebibliography}
